# Artificial Intelligence Specialization in the European Union: Underexplored Role of the Periphery at NUTS-3 Level

Víctor Herrero-Solana
*SCImago-UGR, Unit for Computational Humanities and Social Sciences (U^CHASS)*
University of Granada, Spain
victorhs@ugr.es
ORCID: [0000-0003-1142-5074]

Carmen Gálvez
*Unit for Computational Humanities and Social Sciences (U^CHASS)*
University of Granada, Spain
cgalvez@ugr.es
ORCID: [0000-0001-7454-1254]

## Statements and Declarations

**Competing Interests:** The authors have no relevant financial or non-financial interests to disclose.

**Funding:** This publication is part of the R&D project «Artificial Intelligence in Europe: Rise or Decline? A Bibliometric/Patentometric Approach and Its Impact on Media and Social Networks (innovIA-EU)» (PID2023-149646NB-I00), funded by the 2023 Call for Knowledge Generation Projects of the Spanish Ministry of Science, Innovation and Universities (MCIN/10.13039/501100011033) and by FEDER "Una manera de hacer Europa".

**Data Availability:** The dataset generated and analysed during the current study is available from the corresponding author on reasonable request. A restricted-access version has been deposited in Zenodo portal: https://doi.org/10.5281/zenodo.18645519

**Author Contributions (CRediT):** Conceptualization: Víctor Herrero-Solana; Data curation: Víctor Herrero-Solana; Formal analysis: Víctor Herrero-Solana, Carmen Gálvez; Funding acquisition: Víctor Herrero-Solana; Investigation: Víctor Herrero-Solana; Methodology: Víctor Herrero-Solana, Carmen Gálvez; Validation: Carmen Gálvez; Visualization: Víctor Herrero-Solana; Writing – original draft: Víctor Herrero-Solana, Carmen Gálvez; Writing – review & editing: Víctor Herrero-Solana, Carmen Gálvez.

## Abstract

This study examines the geographical distribution of Artificial Intelligence (AI) research across European NUTS-3 regions during the period 2015–2024. Using bibliometric data from Clarivate InCites and the Citation Topics classification system, we analyse two hierarchical thematic levels: Electrical Engineering, Electronics & Computer Science (Macro Citation Topic 4) and Artificial Intelligence & Machine Learning (Meso Citation Topic 4.61). Relative Specialization Index (RSI) and Relative Citation Impact (RCI) indicators are calculated for 781 European NUTS-3 regions.

While major metropolitan hubs such as Paris (Île-de-France), Warszawa, and Madrid dominate in absolute publication volume, the results reveal that the highest levels of relative AI specialization are concentrated in peripheral regions, particularly in Eastern Europe and Spain. Granada and Vilniaus apskritis stand out as regions combining high specialization with strong citation visibility. The analysis further suggests a weak relationship between regional specialization and citation impact, revealing multiple regional profiles, including highly specialized regions with limited citation visibility, highly visible regions with comparatively low specialization, and diversified scientific systems combining moderate specialization with strong citation impact. In particular, Fyn (Denmark) emerges as an extreme case of very high citation impact despite relatively low specialization.

The findings constitute a contribution to the spatial scientometrics of AI by demonstrating that scientific specialization, citation visibility, and technological leadership follow partially distinct territorial dynamics within the European research landscape. More broadly, the results suggest that AI may function as a domain through which peripheral regions develop specialized scientific niches within the European research system, although achieving international visibility appears to depend on factors extending beyond publication concentration alone.

## 1 Introduction

Over the past decade, Artificial Intelligence (AI) has become a central driver of scientific, technological, and economic change, attracting growing attention from policymakers, industry, and the research community (Arranz et al., 2023). In parallel, the rapid expansion of AI-related scientific publications has stimulated a substantial body of bibliometric and scientometric research that seeks to characterize the evolution, structure, and impact of AI research at the global level (Hajkowicz et al., 2023). Recent bibliometric studies have employed various approaches to map the AI research landscape, from co-citation networks and keyword analysis to institutional collaboration patterns (Bajpai et al., 2025; Paulo and Brambilla, 2025). These studies typically map the worldwide landscape of AI, identifying leading countries, institutions, and collaboration networks, and documenting the sustained growth and diversification of AI output across multiple scientific domains.

A parallel strand of research has explored the spatial distribution of scientific activity more broadly. Spatial scientometrics, as reviewed by Frenken and Hoekman (2014), encompasses a range of methods for analysing the geographical dimension of knowledge production. Grossetti et al. (2014) demonstrated, using a multilevel analysis of publications from 1987 to 2007, that scientific activity has undergone a process of geographical deconcentration at the global level, with cities outside the traditional centres gaining an increasing share of world scientific output. Maisonobe et al. (2017) extended this analysis to scientific visibility, showing that a deconcentration process is also observable when considering the geography of citations, although at a slower pace. These findings suggest that the geography of science is more dynamic than commonly assumed, with peripheral locations progressively gaining ground. However, most spatial scientometric studies have focused on general scientific output or on specific disciplines other than AI.

The geography of AI research specifically has attracted growing attention in recent years. Schmallenbach et al. (2024) analysed nearly 400,000 AI publications in the life sciences from 2000 to 2022, revealing stark geographical concentration: while Asia leads in total publications, North America and Europe contribute the most impactful work, generating up to 50% more citations than other regions. Xiao and Boschma (2023) examined how a regional knowledge base in information and communication technologies (ICTs) influences the emergence of AI technologies in European regions, finding that ICTs play a critically enabling role for regional diversification into AI, especially for catching-up regions. These studies, however, mainly examine AI geography at national or NUTS-2 regional scales, leaving finer-grained territorial dynamics at the NUTS-3 level largely unexplored.

Despite this growing body of research, the subnational geography of AI research remains insufficiently explored. While some studies have examined AI publication patterns at the national level within Europe (Frankowska & Pawlik, 2022), and others have analysed AI-related innovation at broader regional scales, most analyses overlook the substantial heterogeneity that exists within countries. The European Union, with its complex mosaic of regions characterized by varying levels of economic development, industrial specialization, and research infrastructure, provides an ideal setting for examining the territorial distribution of AI research capabilities.

The Nomenclature of Territorial Units for Statistics (NUTS), developed by Eurostat, offers a hierarchical classification system that enables systematic comparison across regions at different scales. The NUTS classification comprises three nested levels, each defined by population thresholds. NUTS-1 corresponds to major socio-economic regions, typically with populations between 3 and 7 million inhabitants; examples include German *Länder* and French *ZEAT* (zones d'études et d'aménagement du territoire). NUTS-2 represents basic regions for the application of regional policies, with populations ranging from 800,000 to 3 million; these correspond to entities such as Spanish *comunidades autónomas*, French *régions*, or Italian *regioni*. Finally, NUTS-3 identifies small regions for specific diagnoses, with populations between 150,000 and 800,000, typically corresponding to provinces, counties, or departments—for instance, Spanish *provincias*, German *Kreise*, or French *départements*. The current NUTS 2024 classification covers 92 NUTS-1 regions, 242 NUTS-2 regions, and 1,166 NUTS-3 regions across the 27 EU member states. This fine-grained territorial resolution at the NUTS-3 level makes it possible to identify localized centers of research excellence that might be obscured in analyses conducted at the national or NUTS-1 level.

In this study, we propose a regional perspective on scientific production, following the broader argument developed by Hautala (2024), who notes that most existing research has focused on major AI clusters and highly visible technological hubs, while peripheral regions remain comparatively understudied. We argue that analysis at the NUTS-3 level enables the fine-grained identification of peripheral regions showing strong relative specialization in AI research. Rather than focusing on absolute publication volumes, which inherently favor large metropolitan areas with extensive research infrastructure, we employ relative specialization indicators that measure the share of regional research activity devoted to AI compared with broader disciplinary baselines. This approach reveals patterns that remain hidden in volume-based analyses: regions with modest scientific output may nonetheless display strong AI specialization within their national or European context. These findings suggest that the geography of AI research may be more territorially distributed than commonly assumed.

## 2 Material and methods

### 2.1 Data sources

In this study, we analyse scientific production in AI across European NUTS-3 regions during the period 2015–2024. Bibliometric data were retrieved from Clarivate InCites, using the Web of Science Core Collection as the underlying database. For the delineation of the thematic domain, Citation Topics were preferred over traditional subject categories. Citation Topics represent a relatively recent innovation introduced by Clarivate to provide document-level classification of scientific publications. Unlike journal-based subject categories, Citation Topics are algorithmically derived citation clusters developed in collaboration with the Centre for Science and Technology Studies (CWTS) at Leiden University (Potter, 2020). This approach groups papers according to their citation relationships rather than the categories of the journals in which they are published. The resulting hierarchical classification system comprises three levels: 10 macro-topics, 326 meso-topics, and 2,444 micro-topics. This document-level classification provides a finer-grained thematic structure that facilitates more detailed domain analyses, as each article is assigned to a specific topic according to its actual citation patterns rather than inheriting the category of its publication venue (Salvador-Mata et al., 2024). For the present study, two hierarchical levels of thematic aggregation were considered: a general level, Electrical Engineering, Electronics & Computer Science (COMPU), corresponding to Macro Citation Topic 4; and a specific level, Artificial Intelligence & Machine Learning (AI), corresponding to Meso Citation Topic 4.61.

The dataset includes publication and citation counts for each NUTS-3 region across the selected thematic levels. The use of bibliometric indicators to assess scientific specialization at subnational territorial scales has previously been explored by Abramo et al. (2014), who applied specialization indices to Italian NUTS-2 and NUTS-3 regions. Following this territorial perspective, our nested data structure enables the calculation of relative specialization indicators that contextualize AI-related research activity within broader disciplinary and scientific frameworks.

## 2.2 The Activity Index

To measure the relative research effort devoted to AI by each European region, we build upon the Activity Index (AIndx), a commonly employed bibliometric indicator introduced by Frame (1977) and formalised by Schubert and Braun (1986). In the present study, the abbreviation AIndx is preferred over the conventional AI notation in order to avoid confusion with the widely used acronym for Artificial Intelligence. This indicator is mathematically equivalent to the Revealed Comparative Advantage (RCA) index proposed by Balassa (1965) in international trade economics (Rousseau & Yang, 2012). The AIndx mitigates the confounding effects of both the size of the region and the size of the field, enabling meaningful cross-regional comparisons (Rousseau, 2019). As a relative indicator, the Activity Index is particularly suitable for identifying regional specialization patterns independently of absolute publication volumes, although its interpretation requires caution in regions with very small publication counts. In this regard, Rousseau and Yang (2012) noted that the indicator may exhibit counterintuitive behaviour under certain conditions, since changes in activity in unrelated fields or regions can affect the resulting values. The indicator is defined as:

$$\mathrm{AIndx}(\mathrm{r,f}) = \frac{\frac{P(\mathrm{r,f})}{P(\mathrm{r})}}{\frac{P(\mathrm{W,f})}{P(\mathrm{W})}} \quad (1)$$

where P(r,f) denotes the number of publications by region r in field f; P(r) is the total publications by region r; P(W,f) represents the total publications in field f at the reference level (Europe); and P(W) is the total scientific production at the reference level.

## 2.3 The Relative Specialization Index

A well-known limitation of the AIndx is its asymmetric distribution: values range from 0 to 1 for under-specialization but can theoretically extend to very large positive values for specialization (Mansourzadeh et al., 2019; Laursen, 2015). This asymmetry complicates comparative analysis and statistical treatment. To address this limitation, we adopt the Relative Specialization Index (RSI), also known as the Revealed Symmetric Comparative Advantage (RSCA), proposed by Laursen (2015) as a normalized transformation of the AIndx:

$$\mathrm{RSI}(\mathrm{r,f}) = \frac{\mathrm{AIndx}(\mathrm{r,f}) - 1}{\mathrm{AIndx}(\mathrm{r,f}) + 1} \quad (2)$$

The RSI transforms the AIndx into a symmetric measure bounded between −1 and +1, with 0 as the neutral value. This transformation offers several advantages: (1) it provides a symmetric distribution around the neutral point, facilitating interpretation and comparison; (2) it facilitates the use of standard parametric statistical methods; and (3) empirical evidence suggests that RSI values exhibit improved normality in regression residuals compared to untransformed AIndx values (Laursen, 2015).

The RSI has a straightforward interpretation. RSI = 0: The region's share of publications in the field equals the European average (no specialization). RSI > 0: The region is specialized in the field, displaying a higher-than-average publication profile in that area. Values approaching +1 indicate strong specialization. RSI < 0: The region is under-specialized in the field, displaying a relatively lower publication profile compared to the European baseline. Values approaching −1 indicate strong under-specialization.

## 2.4 Relative Citation Impact

To complement the analysis of research specialization with a measure of citation performance, we also compute a Relative Citation Impact (RCI) indicator for each region in the AI field, following the broader tradition of normalized citation indicators in bibliometrics (Moed et al., 1995):

$$\mathrm{RCI}(\mathrm{r,AI}) = \frac{\frac{C(\mathrm{r,AI})}{P(\mathrm{r,AI})}}{\frac{C(\mathrm{W,AI})}{P(\mathrm{W,AI})}} \quad (3)$$

where C denotes citation counts. The indicator compares the average number of citations received per publication in a given region with the corresponding European average in the AI field. An RCI value above 1.0 indicates that AI publications from the region receive more citations per publication than the European average, suggesting higher citation visibility.

# 3 Results

A first approximation to the dataset is presented in Fig. 1, which displays a scatter plot of the relationship between documents published and citations received. The size and color of the circles correspond to the RSI indicator. A small group of regions clearly departs from the general distribution. On one hand, we observe the region with the highest production, Warszawa, followed at a distance by Paris and Madrid. Granada exhibits a publication volume comparable to Paris and Madrid, but with a substantially higher number of citations, the highest in the dataset, reflecting its exceptionally high citation count. The fifth notable element is the Vilniaus region, which, although it does not reach the publication levels of Paris or Madrid, is nevertheless among the regions with the highest citation counts. The remaining regions form a more homogeneous cluster, among which Praha stands out for publication output and Jaén for citation accumulation.

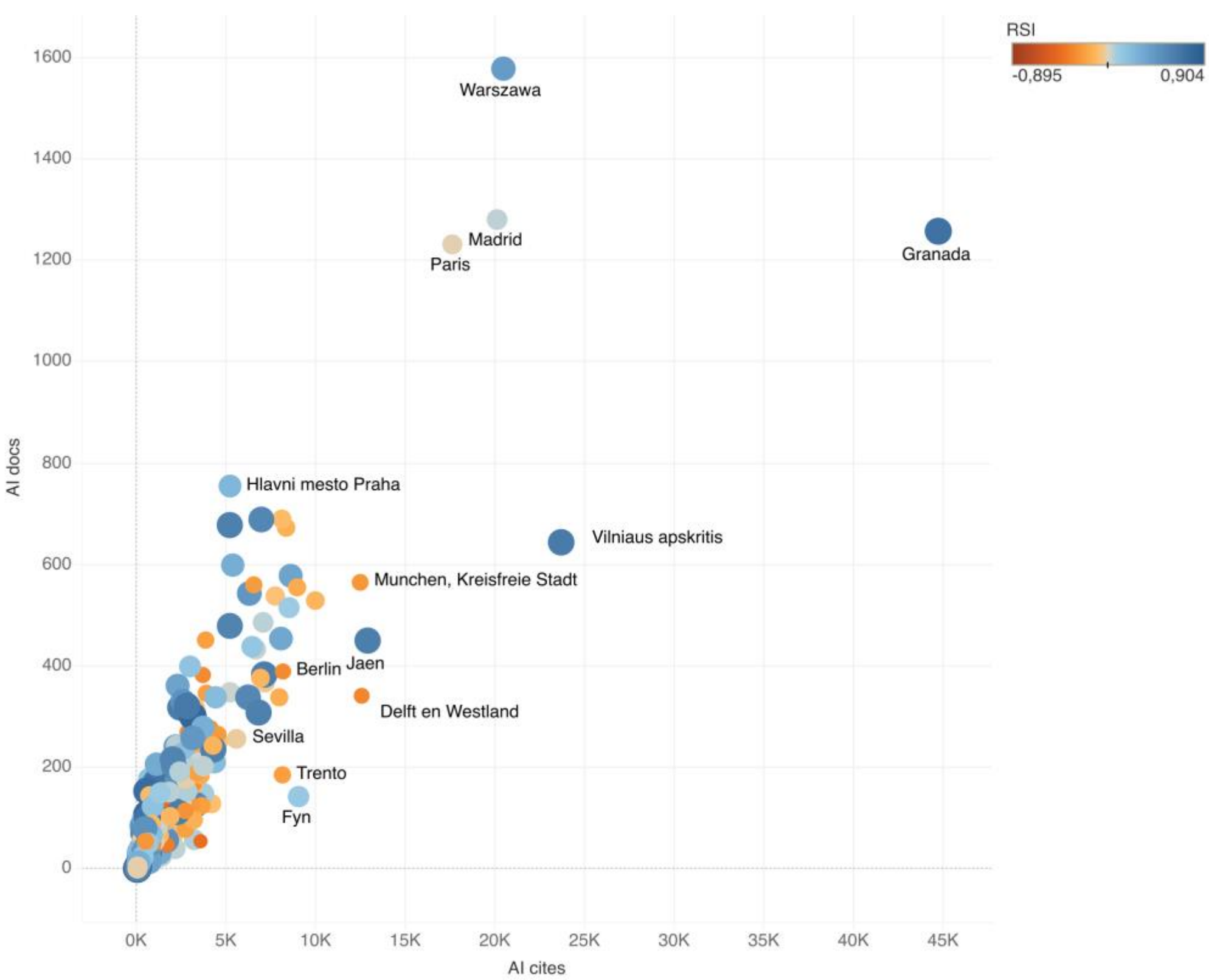

**Fig. 1.** Scatter plot of AI publications and citation counts across European NUTS-3 regions. Circle size and color represent the Relative Specialization Index (RSI).

Table 1 presents the top 20 regions ranked by RSI. To construct this table, regions with fewer than 200 articles in COMPU during the study period were excluded. Regions outside the major European scientific cores clearly dominate the ranking, predominantly from Eastern and Southern Europe. Only one region from the economically central European countries appears in the list: Landkreis Gießen, in Germany. Polish regions are particularly well represented, with several territories, including Białostocki, Krakowski, and Częstochowski, appearing among the most specialized. Spanish regions such as Granada, Jaén, Salamanca, Burgos, Córdoba, Asturias, and Navarra also feature prominently, alongside Czechia (Olomoucký kraj, Jihočeský kraj, and Moravskoslezský kraj), Bulgarian (Burgas and Plovdiv), Slovakian (Banskobystrický kraj), Romanian (Bihor), and Lithuanian (Vilniaus apskritis) regions. Notably, most of these territories do not correspond to the largest European scientific producers in absolute terms. Rather, their position in the ranking reflects a strong thematic orientation toward AI within comparatively smaller regional research systems. This pattern highlights the importance of relative specialization indicators, as analyses based exclusively on publication volume would tend to privilege major metropolitan hubs while overlooking smaller territories with highly concentrated AI research profiles. The predominance of peripheral regions in this ranking is striking and suggests that AI research may constitute a field in which smaller and less diversified regional research systems can nonetheless achieve high levels of relative publication concentration.

This phenomenon can be visualized geographically in Fig. 2, which presents a choropleth map of RSI values across European NUTS-3 regions. The spatial distribution reveals a striking pattern: the highest levels of relative AI specialization, represented by darker blue tones, are concentrated primarily in regions outside the major European scientific cores rather than in the continent's largest and most diversified scientific hubs. Eastern European regions, particularly in Poland, Czechia, Bulgaria, and Romania, display pronounced specialization, as do several Spanish regions distributed across both northern and southern parts of the country. In contrast, many regions belonging to the traditional scientific powerhouses of Western Europe exhibit more moderate RSI values, despite their very high levels of absolute scientific production. This distinction is important because the RSI measures relative thematic concentration rather than total research volume. Consequently, regions with smaller overall scientific systems may nonetheless appear highly specialized when AI-related publications represent a comparatively large share of their scientific output.

| **Region** | **Country** | **RSI** |
|---|---|---|
| Burgas | Bulgaria | 0.820 |
| Olomoucký kraj | Czechia | 0.801 |
| Landkreis Gießen | Germany | 0.751 |
| Granada | Spain | 0.712 |
| Banskobystrický kraj | Slovakia | 0.706 |
| Jihočeský kraj | Czech Rep | 0.702 |
| Krakowski | Poland | 0.692 |
| Częstochowski | Poland | 0.665 |
| Vilniaus apskritis | Lithuania | 0.647 |
| Rzeszowski | Poland | 0.641 |
| Burgos | Spain | 0.631 |
| Navarra | Spain | 0.630 |
| Jaén | Spain | 0.621 |
| Córdoba | Spain | 0.615 |
| Salamanca | Spain | 0.609 |
| Moravskoslezský kraj | Czechia | 0.608 |
| Bihor | Romania | 0.606 |
| Asturias | Spain | 0.585 |
| Białostocki | Poland | 0.579 |
| Plovdiv | Bulgaria | 0.566 |

**Table 1.** Top-20 RSI regions

The map also reveals that high levels of AI specialization are not confined to a single continuous macro-region, but instead appear as dispersed territorial clusters distributed across different parts of Europe. Particularly notable is the concentration of highly specialized regions across Central and Eastern Europe, suggesting that AI may represent a strategic opportunity for regions operating outside the traditional core of European science. A similar pattern can be observed in Spain, where several medium-sized regions achieve high RSI values despite not being among the continent's largest producers in absolute terms. These results reinforce the importance of using relative specialization indicators in spatial scientometric analyses, as approaches based exclusively on publication volume would tend to privilege large metropolitan systems while obscuring smaller territories with highly concentrated AI research profiles

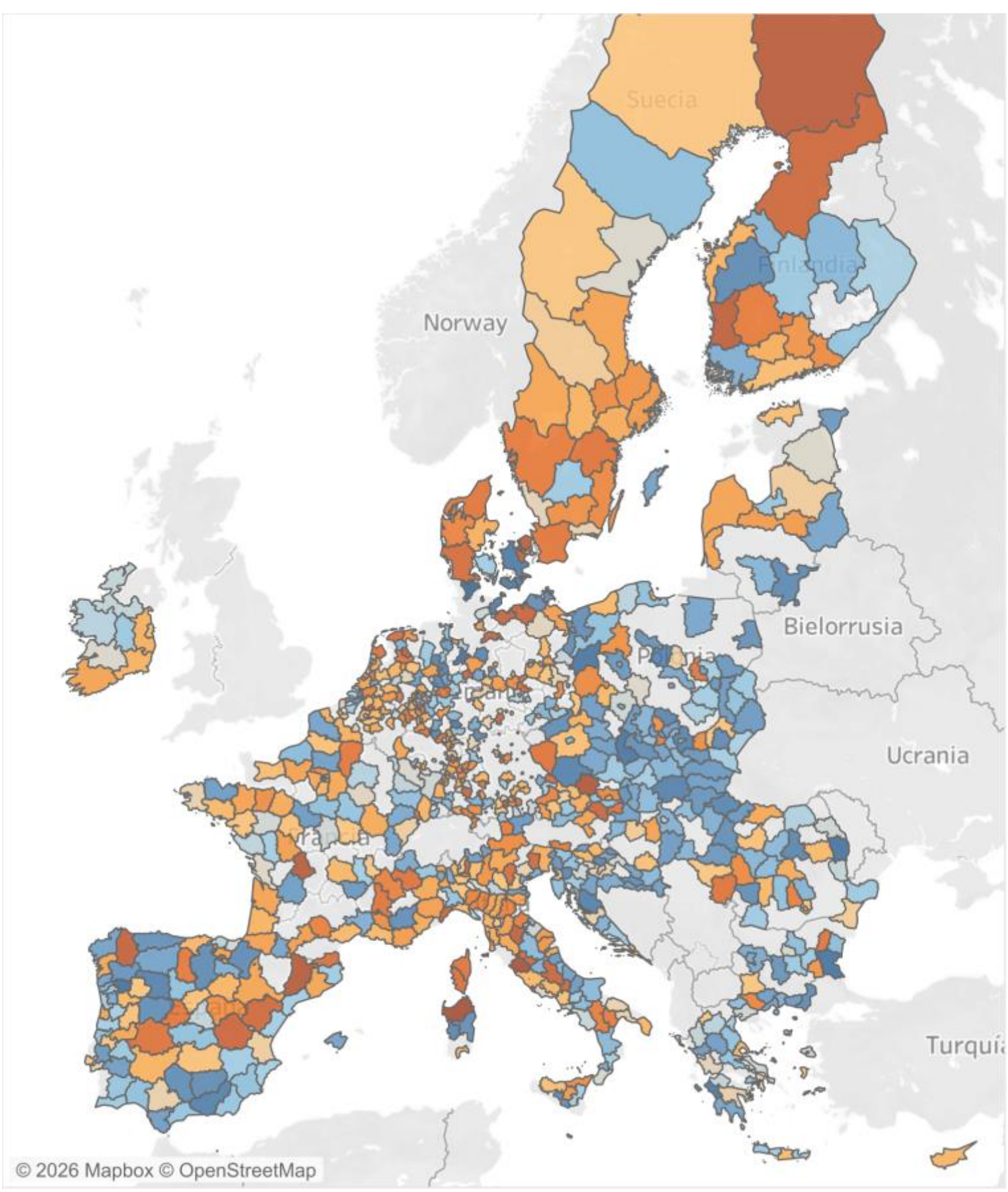

**Fig. 2.** Spatial distribution of Relative Specialization Index (RSI) values across European NUTS-3 regions

In contrast, many regions belonging to the traditional scientific core of Europe, including large parts of France, the United Kingdom, and the Benelux countries, exhibit relatively low or neutral RSI values. This does not imply low AI production in absolute terms, but rather indicates that AI represents a comparatively smaller share of their broader computer science activity. The observed geographical pattern therefore suggests that relative specialization in AI is more territorially distributed than overall scientific production, with several peripheral regions displaying highly concentrated AI research profiles. These findings may indicate that AI constitutes a domain in which smaller regional systems can achieve visible thematic differentiation within the European research landscape.

Regarding the citation impact indicator RCI, Fig. 3 presents it in combination with the RSI, revealing the relationship between specialization and research impact across European regions with at least 100 papers in AI. The scatter plot positions each NUTS-3 region according to its degree of AI specialization (horizontal axis, RSI) and the relative citation impact of its AI publications (vertical axis, RCI). The reference lines at RSI = 0 and RCI = 1.0 divide the space into four quadrants, each representing a distinct regional profile. This visualization allows us to move beyond simple rankings and explore whether regions with strong thematic concentration in AI also achieve high citation visibility, or whether specialization and citation impact follow partially independent trajectories.

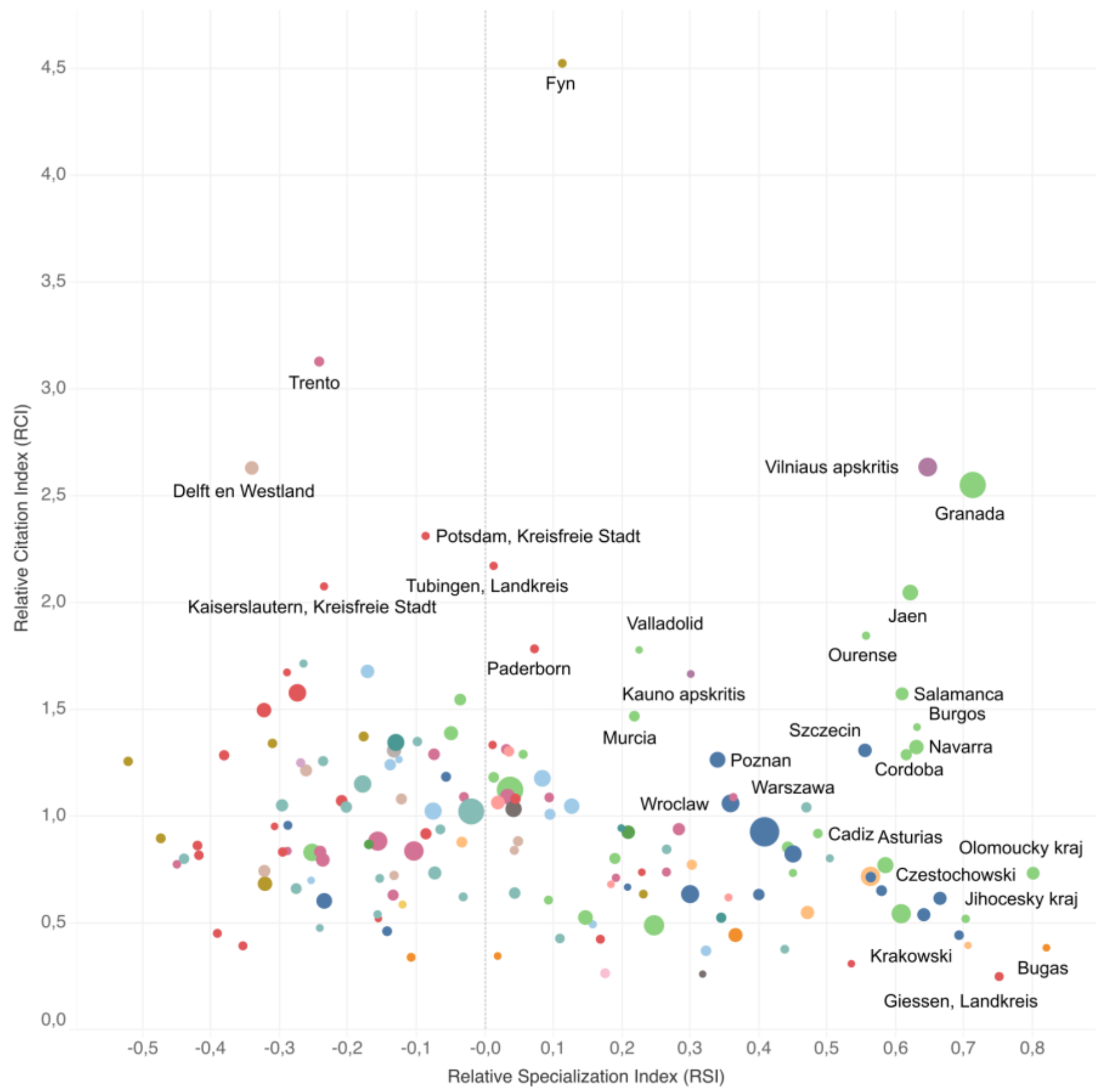

**Fig. 3.** Scatter plot of RSI versus RCI for European NUTS-3 regions with at least 100 AI publications. Reference lines at RSI = 0 and RCI = 1.0 define four quadrants representing distinct regional profiles

Fig. 3 reveals a significant observation: the visual distribution suggests a weak relationship between regional specialization in AI and citation impact. The scatter plot shows substantial vertical dispersion across nearly the entire RSI range, indicating that strong specialization in AI does not necessarily translate into higher citation visibility. This pattern makes it possible to identify several distinct regional profiles within the European AI landscape. The upper-right quadrant (RSI > 0, RCI > 1) contains regions combining both high specialization and above-average citation impact. Spanish regions are particularly prominent in this quadrant, with Granada (RSI = 0.71, RCI = 2.55), Jaén (RSI = 0.62, RCI = 2.05), Salamanca, and Navarra all occupying favorable positions. Vilniaus apskritis (Lithuania) also stands out within this group. These territories appear to have developed AI research profiles that are simultaneously highly specialized and internationally visible in citation terms.

In contrast, the lower-right quadrant (RSI > 0, RCI < 1) groups regions characterized by strong specialization but comparatively limited citation visibility. Several Eastern European territories, including Burgas (Bulgaria) and multiple Polish regions, are concentrated in this area of the plot. These regions devote a substantial share of their scientific activity to AI, yet their citation performance remains below the European average. This pattern may be potentially linked to structural factors such as lower integration into international collaboration networks, reduced institutional visibility, language effects, or more limited access to high-impact publication venues. The concentration of Eastern European regions in this quadrant is broadly consistent with the findings of Fiala and Willett (2015), who observed that computer science research in Eastern Europe expanded substantially after 1989 while continuing to face difficulties in achieving proportional citation impact. The persistence of this pattern in the AI domain suggests that some structural asymmetries within the European scientific system may continue to influence citation visibility across regions.

The upper-left quadrant (RSI < 0, RCI > 1) presents a different profile, characterized by high citation impact despite low specialization. Fyn (Denmark) appears as the most extreme case in the dataset, reaching an RCI above 4.0 while maintaining a negative RSI value. However, this result should be interpreted cautiously, as relatively small publication volumes may amplify citation-based indicators. Trento (Italy) and Delft en Westland (Netherlands) also occupy this quadrant, although with less extreme values. These regions produce a comparatively limited share of AI publications within their broader COMPU output, yet the AI publications they do produce achieve substantial citation visibility. This configuration may reflect selective participation in highly visible international collaborations, concentration in niche research areas, or strategic prioritization of fewer but more influential publications.

The Nordic regions deserve separate consideration because they display particularly heterogeneous patterns in the RSI-RCI space. Although Nordic countries generally rank among the strongest scientific systems in Europe, their NUTS-3 regions are distributed across multiple quadrants rather than forming a coherent regional profile. Some Danish and Finnish regions cluster around neutral RSI values with moderate citation impact, whereas others combine low specialization with very high visibility. This heterogeneity highlights one of the main advantages of NUTS-3 analysis: national-level averages may obscure substantial territorial variation in both AI specialization and citation performance.

German regions also present an interesting pattern. Despite Germany's strong overall scientific capacity, most German NUTS-3 regions are located in the left half of the plot, indicating that AI represents a comparatively smaller share of their broader engineering and computer science activity. At the same time, several German regions, including Potsdam and Tübingen, achieve relatively high RCI values despite negative RSI scores. This suggests the presence of highly visible AI research embedded within broader and more diversified scientific portfolios, rather than within strongly AI-specialized regional systems.

## 4 Discussion

The results suggest that relative specialization in AI is not exclusively concentrated in the major hubs of European science and knowledge production. Instead, high levels of AI specialization frequently emerge in regions located outside the traditional scientific core of Europe. In this sense, AI research appears to provide an opportunity for peripheral regions to develop distinctive thematic profiles within the broader European research landscape. This finding aligns with the observations of Fiala and Willett (2015), who identified a strong orientation toward AI and computer science in several Eastern European countries, particularly Poland and the Baltic states, during the decades following the fall of the Berlin Wall. However, those authors also noted that this increase in production was not accompanied by proportional citation impact, a pattern that remains visible in our results.

At the same time, high RSI values in peripheral regions should not be interpreted as evidence of absolute scientific dominance, but rather as indicators of thematic concentration within comparatively smaller regional scientific systems. Regions with less diversified scientific portfolios may achieve high relative specialization when AI-related activity constitutes a substantial share of their total computer science output. This interpretation is consistent with the literature on regional diversification and niche formation, which argues that peripheral regions may develop competitive specialization strategies by concentrating resources in selected emerging domains (Boschma, 2017). Our findings therefore suggest that AI may function as a niche specialization domain through which smaller regional systems can increase their scientific visibility within Europe, even when they do not possess the scale or infrastructural capacity of the continent's largest metropolitan centers.

The importance of peripheral regions appears particularly visible in terms of scientific production measured through publications. However, the same pattern does not seem to apply to technological innovation measured through patents. Although bibliometric and patent datasets capture different dimensions of knowledge production and innovation, their combined interpretation offers a broader perspective on the structure of regional AI ecosystems. In the patent domain, technological innovation appears to remain more strongly associated with major metropolitan and innovation hubs. Buarque et al. (2020), for example, showed that a small number of central European regions concentrate a very large share of AI patents. Xiao and Boschma (2023) reached similar conclusions regarding the territorial concentration of AI technological capabilities, while Santos et al. (2025) found that spillover effects associated with AI funding are considerably stronger in central regions than in peripheral territories. These findings are also broadly consistent with Balland et al. (2020), who argued that complex and frontier technological activities tend to concentrate disproportionately in large metropolitan ecosystems characterized by dense knowledge networks and strong institutional connectivity.

Cicerone et al. (2022) provide an interesting complementary perspective through their analysis of AI and green technologies at the NUTS-3 level. Although they do not identify a clear spillover effect, their work suggests that peripheral regions may participate actively in scientific production without necessarily becoming major technological innovation hubs. Research following this line for the Spanish case reveals a particularly clear contrast: Spain performs strongly in scientific AI publications but occupies a comparatively modest position in AI patenting (Herrero & Jürgens, 2025). Extending this perspective to the broader European context, preliminary evidence identifies Paris (Île-de-France) as one of the continent's most important hubs for AI patenting and technological innovation. This pattern reinforces the distinction between scientific specialization and technological leadership: relative specialization in AI research may be territorially distributed, while technological innovation remains substantially more concentrated.

The disconnect between specialization and citation impact observed in our results also deserves particular attention. While several Eastern European regions demonstrate strong commitment to AI research in terms of relative publication concentration, their citation visibility frequently remains below the European average. This pattern may be potentially linked to structural factors such as lower integration into international co-authorship networks, limited institutional visibility, historical path dependencies in research systems, or unequal access to highly visible publication venues. By contrast, regions such as Fyn (Denmark) or Delft en Westland (Netherlands) achieve exceptionally high citation impact despite relatively modest specialization levels. However, these cases should be interpreted cautiously, as citation-based indicators may become volatile in regions with comparatively small publication volumes. Even so, these patterns potentially suggest that selective participation in highly visible international collaborations, concentration in niche research areas, and strategic prioritization of fewer but highly influential publications may play a more important role in global citation visibility than publication volume alone.

The Spanish case is particularly illustrative within the European context. Several Spanish NUTS-3 regions, especially Granada, Jaén, Salamanca, and Navarra, combine high specialization with above-average citation impact, occupying favorable positions in the RSI-RCI space. This suggests that sustained regional investment in AI research, when accompanied by active participation in international scientific networks, may generate internationally visible research profiles even outside the traditional metropolitan cores of European science. In this respect, the Spanish pattern contrasts with the Eastern European case, where high specialization has not yet translated into comparable citation visibility. These differences raise important questions concerning the role of national research policies, funding mechanisms, institutional strategies, and international collaboration frameworks in mediating the relationship between specialization and scientific impact.

Our findings also carry implications for the debate surrounding Smart Specialization Strategies (S3) within the European Union. The territorial patterns observed here are consistent with the broader S3 framework proposed by Foray (2015), which emphasizes the importance of regionally differentiated innovation trajectories based on existing capabilities and selective specialization. The evidence presented in this study suggests that AI specialization does not emerge exclusively in highly resourced metropolitan systems, but may also develop in peripheral territories through concentrated thematic orientation and institutional strategic choices. At the same time, our results caution against interpreting high relative specialization as a sufficient condition for scientific excellence or technological leadership. The citation impact dimension reveals that additional investments in internationalization, infrastructure, talent attraction, and network integration remain essential for transforming thematic specialization into globally visible and technologically competitive research ecosystems.

More broadly, the results are also consistent with recent discussions concerning territorial inequalities and uneven development across Europe. Rodríguez-Pose (2018) argued that many peripheral regions have become increasingly disconnected from the main dynamics of economic and technological growth. Our findings suggest a more nuanced picture in the AI domain: although peripheral regions rarely dominate in absolute scientific output or technological innovation, several appear capable of constructing visible and highly specialized scientific niches within the European research system. AI may therefore represent not only a frontier technological field, but also a potential pathway for territorial differentiation within Europe's evolving knowledge economy.

## 5 Conclusions

This study provides a large-scale bibliometric analysis of AI research specialization across 781 European NUTS-3 regions for the period 2015–2024. By employing the Relative Specialization Index (RSI) and the Relative Citation Impact (RCI) at this fine-grained territorial level, we identify spatial patterns that remain largely invisible in national-level analyses. The results reveal a complex and differentiated geography of AI research within Europe, in which scientific specialization, citation visibility, and technological innovation follow partially distinct territorial dynamics.

First, AI research specialization in Europe is geographically distributed in a way that challenges conventional assumptions regarding the geography of innovation. Regions outside the traditional scientific core of Europe, particularly in Eastern Europe and Spain, exhibit the highest levels of relative specialization, whereas many established scientific centers display neutral or negative RSI values. These findings suggest that AI constitutes a domain in which peripheral regions may develop increasingly specialized scientific profiles within the European research system. At the same time, high RSI values should not be interpreted as indicators of absolute scientific dominance, but rather as expressions of thematic concentration within comparatively smaller and less diversified regional scientific portfolios.

Second, the relationship between AI specialization and citation visibility appears relatively weak. Strong thematic concentration in AI does not necessarily translate into internationally visible research. The regional profiles identified in the RSI–RCI space reveal multiple territorial configurations, including highly specialized regions with limited citation visibility, highly visible regions with comparatively low specialization, and diversified scientific systems in which AI represents only a limited share of broader research activity. These findings suggest that specialization and citation impact are shaped by partially independent mechanisms, potentially associated with factors such as international collaboration networks, institutional visibility, and regional research capacity.

Third, the contrast between scientific production and technological innovation in AI deserves particular attention. While peripheral regions frequently stand out in terms of relative scientific specialization, available evidence suggests that AI patenting and technological innovation remain substantially more concentrated in a limited number of major metropolitan hubs. Although bibliometric and patent datasets capture different dimensions of knowledge production and innovation, their combined interpretation provides a broader understanding of the territorial structure of European AI ecosystems. In this respect, the results suggest that relative scientific specialization in AI does not automatically translate into technological leadership or innovation capacity.

Several limitations should nevertheless be acknowledged. The analysis relies on Web of Science data accessed through the InCites platform, which may underrepresent research published in regional or non-English-language venues. The NUTS-3 classification, while offering high territorial granularity, assigns publications according to institutional addresses and may therefore not fully capture the geography of collaborative or multi-sited research networks. Furthermore, although Citation Topics provide a sophisticated document-level classification system, the methodology remains relatively recent and its long-term stability has yet to be fully evaluated.

Future research could extend this work in several directions. First, integrating patent data would allow a more systematic examination of the relationship between AI scientific specialization and technological innovation at the regional level. Second, analysing international co-authorship networks may help explain the observed differences between specialization and citation visibility across regions. Third, longitudinal analyses could track the temporal evolution of regional specialization profiles in order to evaluate whether peripheral regions are converging toward or diverging from the dominant scientific centers of Europe. Such approaches would also make it possible to assess the effects of European policy instruments, including Structural Funds and Horizon Europe programmes, on the development of regional AI research capacity.

More broadly, the findings are consistent with recent debates on regional diversification, niche formation, and Smart Specialization Strategies (S3) within the European Union. The results suggest that AI may function as a strategic niche domain through which smaller regional systems attempt to increase their scientific visibility and differentiation within the European knowledge economy. However, the persistence of strong metropolitan concentration in technological innovation also indicates that specialization alone is insufficient to overcome structural territorial inequalities in European research and innovation systems.